\documentclass[twocolumn,english,notitlepage,superscriptaddress,nofootinbib]{revtex4-2}
\usepackage[T1]{fontenc}
\usepackage[latin9]{inputenc}
\usepackage{float}
\usepackage{amsmath}
\usepackage{amssymb}
\usepackage{graphicx}
\usepackage{wasysym}
 \usepackage{color}
\makeatletter

\newcommand{\affone}{Department of Physics and Astronomy, University College London, Gower Street, WC1E 6BT London, United Kingdom.}

\makeatother

\begin{document}
\title{Sensing directional noise baths in levitated optomechanics}
\author{J. M.H.  Gosling} 
\affiliation{\affone} 
\author{A. Pontin}
\email{a.pontin@ucl.ac.uk}
\affiliation{\affone}
\author{J. H. Iacoponi}
\affiliation{\affone}
\author{ P.F.  Barker}
\affiliation{\affone}
\author{T.S. Monteiro}
\email{t.monteiro@ucl.ac.uk}
\affiliation{\affone}

\begin{abstract}
Optomechanical devices are being harnessed as sensors of ultraweak forces for applications ranging from inertial sensing  to the search for the elusive dark matter.  For the latter,  there is a  focus on detection of either higher energy single recoils or ultralight,  narrowband sources; a directional signal  is expected.   However,  the possibility of searching for a stochastic stream of weak  impulses,  or more generally a directional  broadband signal,  need not be excluded; with this and other applications in mind,   we apply Gaussian white noise impulses  with a well defined direction $\Psi$ to a levitated nanosphere trapped and 3D cooled in an optical tweezer.  We find that cross-correlation power spectra offer a calibration-free distinctive signature of the presence of a directional broadband force and its  orientation quadrant,  unlike normal power spectral densities (PSDs). 
We obtain excellent agreement between theoretical and experimental results.  With calibration we are able to  measure the  angle $\Psi$,   akin to a force compass in a plane.  We discuss prospects for extending this technique into quantum regime and compare the expected behaviour of quantum baths and classical baths.
\end{abstract}
\maketitle


\section{Introduction}
Experiments on levitated nanoparticles,  cooled and controlled in cavities or by active feedback~\cite{millen2020optomechanics},  are mainly aimed at  two important goals. The first is the experimental realisation of quantum coherence in systems of mesoscopic or macroscopic size.  The second is the ultra-sensitive detection of forces and accelerations,  with applications ranging from fundamental physics  \cite{Geraci2010,Arvanitaki2013,Carney2021a,Carney2021b} to real world applications.

The two goals are by no means exclusive.  Indeed, combining them for quantum force sensing is itself an overarching aim.  Recent progress has been remarkable: cooling to  quantum (or near) ground state of a single mode of the motion  has been achieved,  either by quantum control~\cite{Magrini2021real,Novotny2021quantum}  or via the optical mode of a cavity~\cite{delic2020cooling}.  Multimode cooling and dynamics is of increasing interest \cite{MTTM2020,MTTM2021,Pontin2023,Novotny2023,PontinNP2023}.
Sensing of forces as low as the zeptoNewton scale has already been investigated experimentally~\cite{Geraci2016,hebestreit2018sensing,dark_matter_3, liang2023yoctonewton, hempston2017force,timberlake2019static} with levitated nanoparticles.  Multimode dynamics and cooling is of increasing interest \cite{MTTM2020,MTTM2021,Pontin2023,Novotny2023,PontinNP2023}.
 Within this second context,  optomechanics based sensing studies have focussed on three distinct types of forces \cite{millen2020optomechanics}: constant forces, harmonic forces and isolated impulses.  
 
\begin{figure}[ht!]
{\includegraphics[width=3.3in]{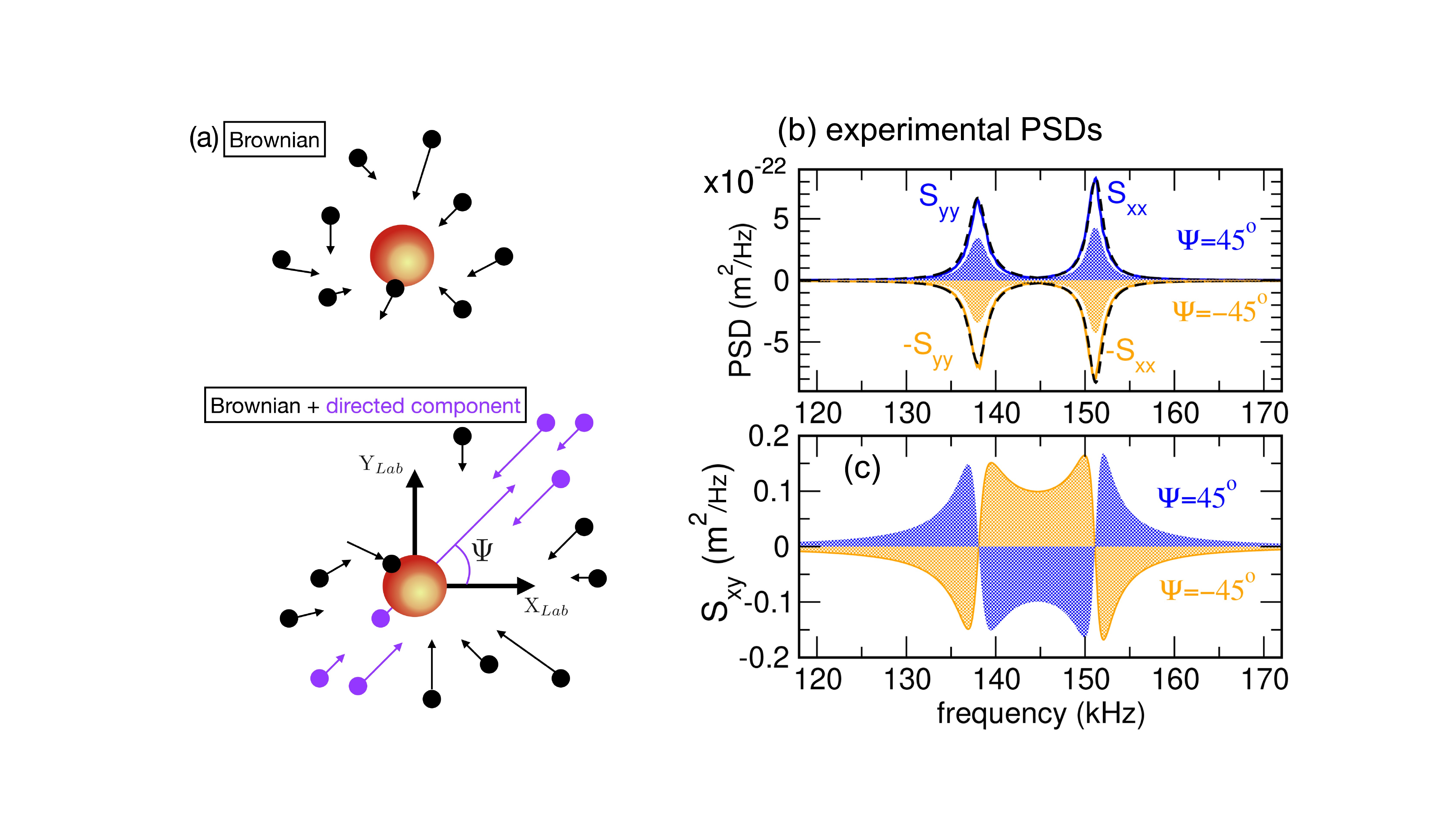}} \caption{ \textbf{(a)} A levitated nanoparticle is  exposed to stochastic forces $ {\bf f}^{th}$ such as  Brownian thermal motion,  uncorrelated in any plane e.g.  $\langle f^{th}_x(t)  f^{th}_y(t') \rangle=0$ (upper image).   Consider detection of a stream of stochastic impulses,  too weak to be individually detected but from fixed orientation $\Psi$ ($45^\circ$ is illustrated).  \textbf{(b)} By applying a directed stochastic force,   we show that with normal PSDs (solid lines)  one must first subtract the thermal component;  and they are  insensitive to  quadrant:  $\pm \Psi$  are indistinguishable.   (Hatched areas show the PSDs without directed noise; solid lines show experimental data with directed noise; dashed lines use  Eq.\ref{PSDs} (for $\beta\simeq \sqrt{2}$, see text).
\textbf{(c)}  Cross-correlation spectra, $S_{xy}(\omega)$,  in contrast,  are shown to offer a calibration-free signature of a directed force and its of orientation quadrant.}
\label{Fig1}
\end{figure}

Here we consider a fourth possibility.   Stochastic forces of microscopic origin are ubiquitous in nature.  Nanoparticles experience {\em thermal} forces ${\bf f}^{th}(t)$,  such as collisions with  surrounding gas molecules; or due to photon recoils.  These represent stochastic noise baths  and are Markovian,  so $ \langle f^{th}_{x,y}(t)  f^{th}_{x,y}(t') \rangle \propto  \delta(t-t')$.   To date it it has been 
 assumed that  $\langle f^{th}_x(t)  f^{th}_y(t') \rangle=0$, i.e.   the corresponding impulses along $x,y$ (and $z$) are uncorrelated and these baths are considered   purely deleterious sources of heating.  However,  if  a subset of these impulses are randomly timed,  but with a preferred orientation,  this assumption may not be justified.
 
  We propose  to detect anisotropies -specifically directionality- via the cross-correlation spectrum    $S_{xy} (\omega)= \frac{1}{2} \left(  \langle [\hat{x}]^\dagger \hat{y} \rangle + \langle [ \hat{y}]^\dagger \hat{x}\rangle \right)$.
 For example,  we consider an incident beam of microscopic particles from a particular direction $\Psi$  in the $x-y$ plane: even if the stream of weak stochastic impulses  are individually undetectable,  integrated over a macroscopic timescale,   the cumulative effect of this white-noise force can generate a measurable, steady-state correlation spectrum $S_{xy}$.  
 
 \begin{figure*}[ht]
{\includegraphics[width=6.in]{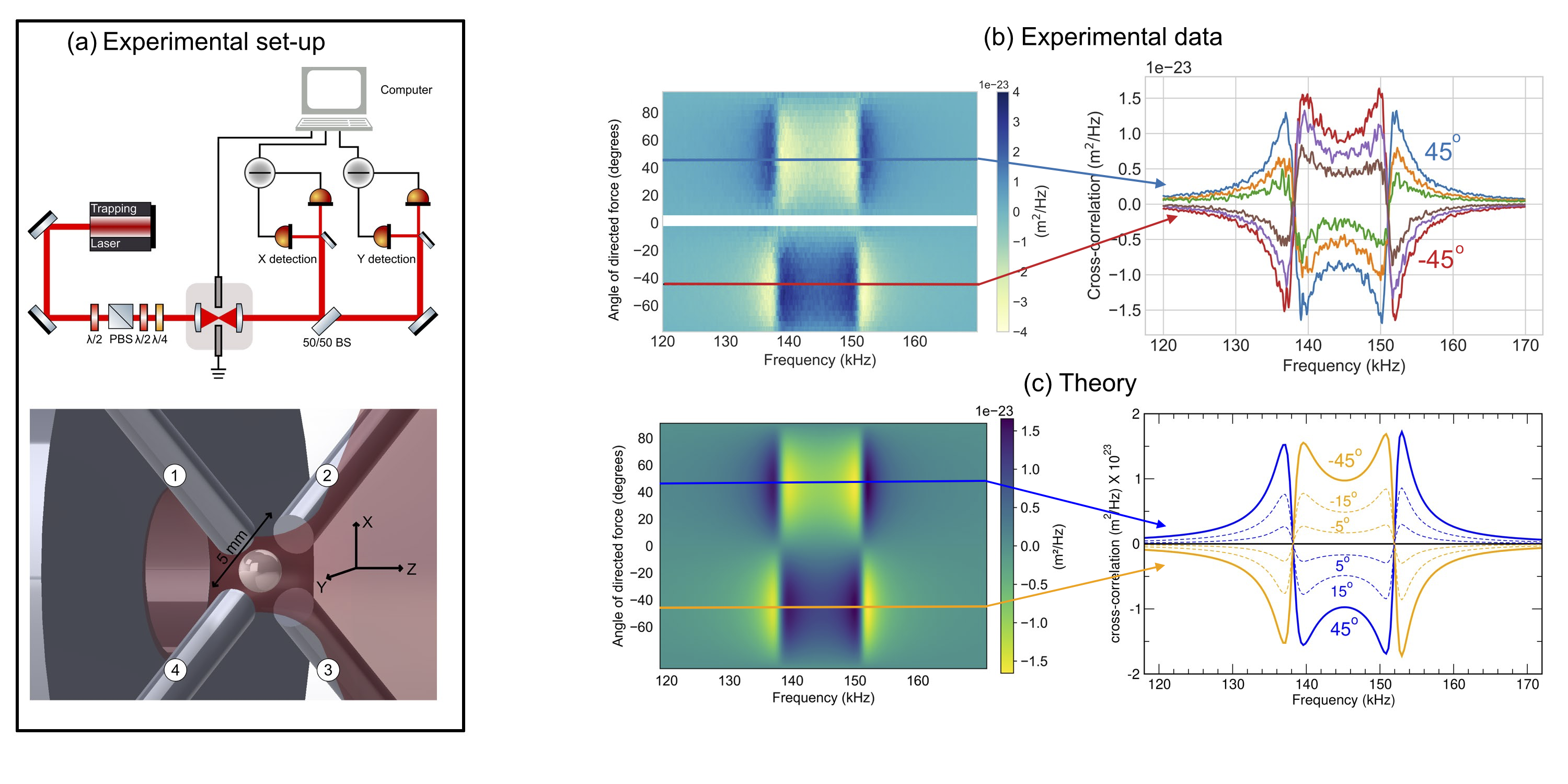}} \caption{\textbf{(a)} Simplified layout of the experimental setup (upper) and schematic view of the trapped nanoparticle with the geometrical configuration of the 4 electrodes exploited for injecting the directional force (lower panel; NB: $R\approx 70$ nm nanosphere image is not to scale).  \textbf{(b)} Experimental data showing the cross-correlation for six different orientations (right panel) and a 2D map of the cross-spectral density $S_{xy}(\omega)$ obtained from a detailed  sweep of $\Psi$  (left panel).  Even without any calibration,  the traces have the `smoking gun' characteristic shape indicating the presence of a directed force; and the flip in sign  that discriminates between orientation quadrants.  \textbf{(c)} Shows theoretical simulation,  with the  analytical expression Eq.\ref{SxyAn},  in excellent qualitative and quantitative agreement with experiment,  showing that we may also measure arbitrary values of $\Psi$ to within under a  degree.  For the experiment $\beta \simeq \sqrt{2}$. }
\label{Fig2}
\end{figure*}

We experimentally demonstrate the effect by applying an electrically generated stochastic force at  orientation $\Psi$ to the $x-y$ axes of a  nanosphere,  trapped in an optical tweezer and 3D cooled by feedback control.  The levitated particle is naturally charged,  but with one to a few elemental charges.  Hence the forces are microscopic-  on the attoNewton scale.  We show the 
$S_{xy}$ signal provides an unambiguous and  calibration-free signature  of the  presence of the directional stochastic noise,  but also  its orientation quadrant,  as illustrated  in Fig.~\ref{Fig1}.  
 
 Applications of detection of such thermal-bath anisotropies may include detection of weak gas currents or temperature anisotropies \cite{Millen2014}.   The exquisite force sensitivity of levitated nano and micro-spheres in high vacuum  offers a promising new approach  in the search for dark matter \cite{Carney2021a,Carney2021b},  for detectable single impacts and narrowband sources.  But the possibility of detecting  a directional stochastic source via the resulting cross-correlations has not yet been considered.  
 The work opens the way to applications in quantum regime and we discuss the signature of a quantum bath.  In our present set-up and cooling,  easily achievable pressures of $10^{-6}-10^{-7}$ mbar cool the motion down to occupancies  $n_{x,y}\sim 10$ quanta.  However, the present sensing technique is demonstrated at  higher pressures.  Displacements  on the scale of quantum zero point motion ($\sim 10^{-12}$m here) may be detected from the phase of the scattered light,  but independent detection of $x,y$ transverse to the axial tweezer direction,   in order to generate quantum $S_{xy}$ spectra,  requires further sensitivity enhancements of the split-detection scheme employed here,
by about 1-2 orders of magnitude.  In addition,  the artificially generated noise and other measurement errors reduce sensitivity.  But this is not a fundamental limit.
 
 In Section II below we introduce the physical model and cross correlation spectra,  while in Sec. III we introduce the experiment.  In Sec.IV we discuss our results.   In Section V we briefly consider quantum regimes and quantum noise baths,  using as an example the quantum optical shot noise in a cavity.   Finally we conclude.

\section{Classical noise model}
We consider a spherical nanoparticle trapped in an optical tweezer.  Its centre of mass motion corresponds
to a set of harmonic oscillators of frequencies $\omega_u$ with  $u\equiv x,y,z$,  under-damped by collisions with background gas that also drive the particle motion with a stochastic force $f_{\text{th}}$.  The nanoparticle is  driven by a feedback force to pre-cool and stabilize its motion.  In addition, we consider a directional force $f_{\text{dir}}$. The equations of motion can then be written as
\begin{equation}\label{eq1}
  \ddot{u}+\gamma_u \dot{u}+\omega_u^2 u=(f_{\text{th},u}(t)+f_{\text{dir},u}(t)+f_{\text{fb},u}(t))/m
\end{equation}

\noindent where $u\equiv x,y,z$ and $\gamma$ is the gas damping~\cite{Epstein1924on,Cavalleri2010gas} while $m$ is the particle mass.  For a spherical particle,  normally gas damping rates are isotropic $\gamma_u\equiv \gamma$. 
 The feedback force  $f_{\text{fb,u}}(t)$ , which for the cold-damping scheme corresponds to a force proportional to the measured velocity $v(t)$.  The  equations can readily be solved in the frequency domain.  In particular,  the  feedback force is then
 approximated $f_{\text{fb},u}(\omega)/m= {\text H}_u(\omega)v(\omega) \approx -i\omega \Gamma^{(u)}_{eff} x(\omega)$ hence: 
\begin{equation}\label{eq2}
u (\omega)=\chi_u(\omega) (f_{\text{th},u}(\omega) +f_{\text{dir},u}(\omega)).
\end{equation}

Hence the transfer function FT$[{\text H}(t)] ={\text H}(\omega)$ was replaced by an effective constant damping
allowing a simpler analysis,  but noting that  the full stochastic numerics can simulate the experimental transfer function.
In this case we obtain the  effective susceptibility $\chi_u=[m(\omega_u^2-\omega^2+i \omega \Gamma_u)]^{-1}$ where we take $\Gamma_u= \gamma+ \Gamma^{(u)}_{eff}$ .  For simplicity we also  neglected the effect of the imprecision noise on the feedback scheme since it is not relevant at the pressure the experiment is carried out at~\cite{Penny2021performance}.

The power spectral densities (PSDs) may be obtained once the properties of the stochastic force noises are defined.  For thermal noise we have zero mean and correlation $\langle f_{\text{th},u}(t) f_{\text{th}, u'}(t') \rangle=2k_B T m \gamma \,\delta(t-t')\delta_{u,u'}$ where $u, u'\equiv x,y,z$, with an isotropic power spectrum $S_{\text{th},u}\equiv S_{\text{ff}}^{\text{th}}$.

 We restrict our analysis to the nanoparticle motion in the 2D  tweezer polarization plane identified by the $x$ and $y$ axes in Fig.~\ref{Fig2}; the $z$  motion remains largely decoupled due to the significant difference in the trap frequencies.  However all calculations and numerics are fully 3D.
 
{\em Directional forces}: For the directional force noises,  we assume zero mean and correlation functions $\langle f_{\text{dir},x}(t)f_{\text{dir},x}(t')\rangle=S_{\text{ff}}^{dir} \cos^2 \Psi \delta(t-t')$, $\langle f_{\text{dir},y}(t)f_{\text{dir},y}(t')\rangle=S_{\text{ff}}^{dir} \sin^2\Psi \delta(t-t')$, and finally a cross correlation $\langle f_{\text{dir},x}(t)f_{\text{dir},y}(t')\rangle=S_{\text{ff}}^{dir} \cos\Psi\sin\Psi \delta(t-t')$. Here, the angle $\Psi$ represents the direction of the stochastic force.  It is convenient to quantify the relative magnitude of the directed and background thermal forces,  especially for the case where both have white noise spectra. Thus we introduce:
\begin{equation}\label{beta}
\beta^2 = S_{\text{ff}}^{dir}/S_{\text{ff}}^{\text{th}}
\end{equation}
The experimental PSDs are then given by:
\begin{eqnarray}
  S_{xx}(\omega) &\simeq &|\chi_x(\omega)|^2 S_{\text{ff}}^{\text{th}}(1+\beta^2\cos^2 \Psi) \nonumber\\
  S_{yy}(\omega) & \simeq &|\chi_y(\omega)|^2 S_{\text{ff}}^{\text{th}} (1+\beta^2\sin^2 \Psi)\nonumber \\
  S_{xy}(\omega) & \simeq &\text{Re}[\chi_x^*(\omega) \chi_y(\omega)] S_{\text{ff}}^{\text{th}}\beta^2\cos \Psi \sin \Psi.\label{PSDs}
\end{eqnarray}

One readily sees  that the cross correlation has the advantage that it isolates the signal due to the directional force from the thermal noise contribution,  since
 $S_{xy}(\omega)$  depends only on $S_{\text{ff}}^{dir}$ (if sensing an uncontrolled force, $\beta$ is not known a-priori).  
  $\beta \ll 1$ corresponds to the regime where the directional force is very weak compared with the thermal component;  while for $\beta \gg 1$,  the directed force is dominant.  For $\beta=\sqrt{2}$ and $\Psi=\pi/4$ (which gives the maximal cross-correlation),  $\beta^2\cos^2 \Psi=\beta^2\sin^2 \Psi=1$ thus the directed contribution to the PSDs  is equal to the thermal contribution; the experiments presented in Fig.\ref{Fig1} and Fig.\ref{Fig2} correspond to $\beta\simeq 1.4\sim \sqrt{2}$.

\section{ Experimental set-up:}

An overview of the experiment is shown in Fig.~\ref{Fig2}a).  A diode pumped solid state laser of wavelength $\lambda=1064$\,nm illuminates a single aspheric lens of nominal numerical aperture NA$=0.77$ to obtain an optical tweezer.  At its focus, we trap single silica nanoparticles with typical radii in the range $R=60-90$\,nm.  A condenser aspheric lens, NA=$0.55$, re-collimates the beam after the tweezer focus.  Subsequently,  the beam is attenuated and divided in two paths by a non-polarizing beam splitter.  On each path, a split detection scheme allows one to measure the particle motion in the tweezer polarization plane $x$-$y$ along orthogonal directions. The particle motion is cooled in all three center of mass (CoM) directions by a cold damping scheme~\cite{poggio2007feedback,Novotny2021quantum,Magrini2021real,kamba2021recoil} exploiting the Coulomb force.  Thus,  the trapped particles are charged with a number of elementary charges  typically in the range $1-50$
- and this represents an essential requirement for the feedback scheme.

Hence,  the  most effective approach is  to introduce a directional stochastic Coulomb force, ${\bf f}_{\text{elec}}$.   In order for ${\bf f}_{\text{elec}} \equiv {\bf f}_{\text{dir}}$ there are specific requirements for the force components in the $x,y$ directions:  (i) vanishing mean values $\langle f_{\text{elec},x} \rangle =\langle f_{\text{elec},y} \rangle =0$,  (ii) correlation $\langle f_{\text{elec},x}(t)f_{\text{elec},x}(t')\rangle=S_{\text{ff}}^{elec} \cos^2 \Psi \delta(t-t')$ and $\langle f_{\text{elec},y}(t)f_{\text{elec},y}(t')\rangle=S_{\text{ff}}^{elec} \sin^2\Psi \delta(t-t')$,  and finally a cross correlation $\langle f_{\text{elec},x}(t)f_{\text{elec},y}(t')\rangle=S_{\text{ff}}^{elec} \cos\Psi\sin\Psi \delta(t-t')$ (iii) while $S_{\text{ff}}^{elec}=S_{th} \beta^2\equiv 2k_BTm \gamma \beta^2$.

To implement this experimentally we use four rod electrodes laying in a plane parallel to the tweezer polarization plane. This is shown in Fig.~\ref{Fig2}a.  The plane of the electrodes is slightly shifted from the origin which represents the particle mean position. The symmetry of this geometrical configuration allows one to have full control of the electric field components in the $x$-$y$ plane by applying the same voltage signal to 3 electrodes each with the right scaling factor while keeping the fourth grounded. This can be intuitively understood by considering that a voltage $V_s$ on electrode-1, with all other grounded, will give an electric field $\textbf{E}_1=V_s\, (-T_x,-T_y,-T_z)^T$ where $T_i$ represents the inverse of an effective distance.  For symmetry reasons, if the same voltage is applied on electrode-2 the electric field would be $\textbf{E}_2=V_s\,(-T_x,T_y,-T_z)^T$. Thus, the superposition of the two electric fields will have a vanishing component on the $y$ direction. The same approach can be used with electrodes-1 and 4 to obtain an electric field with vanishing $x$ component.

The last step to obtain an arbitrary electric field direction in the $x$-$y$ plane is to control the relative amplitude of the voltage signal delivered to electrodes 1-2 and 1-4. We do this with a Wheatstone bridge which given an input $V_s$ provides two outputs $V_{s1}$, $V_{s2}$ whose ratio fully depends on a variable resistor.  By tuning the bridge one can approximate $V_{s1}\simeq V_s \cos \Psi$ and $V_{s2}\simeq V_s \sin \Psi$.

In the experiment, $V_s$ is a white noise signal, provided by a standard signal generator, with a spectral density of $S_{VV}=1.4\times10^{-8}$\,V$^2$/Hz on a bandwidth of $BW=2.5$\, MHz, after amplification with gain $G=80$, $V_{s1}$ and $V_{s2}$ drive electrode-2 and 4 respectively while electrode-1 is driven with their sum.

Experimentally,  we infer $\Gamma_{eff}^{(x)} \approx 2.1$ kHz,  $\Gamma_{eff}^{(y)} \approx 2.5$ kHz.  To simulate the data in  Figs.\ref{Fig1} and \ref{Fig2} we take pressure  $P= 4. \times 10^{-3}$ mbar.

\section{Results:} 

In Fig.\ref{Fig2} we show that there is excellent  agreement,  for all angles $\Psi$,  between the measured cross-correlations and the expression,
\begin{equation}
 S_{xy}(\omega)\simeq \text{Re}[\chi_x(\omega)\chi_y^*(\omega)] 2k_BTm \gamma \beta^2 \sin \Psi \cos \Psi
 \label{SxyAn}
 \end{equation}
by taking $\beta=1.4$. 

The accuracy of the  value of $\beta=1.4$ for modelling the applied electric stochastic noise in Fig.\ref{Fig2},  may be further tested by considering the normal PSDs.
 The directed force can affect the usual PSDs quantitatively.  If the directed force provides heating,  the final temperature is increased by the factor $(1+\beta^2 \cos^2 \Psi)$ for the $x$ mode,  and $(1+\beta^2 \sin^2 \Psi)$
 for $y$.  In the absence of the directed force,  we measured $T_y=215.7 \pm 0.6$ mK and $T_x=297 \pm 2$ mK.
 In the presence of the directed noise,  for the same pressure,  the temperatures are enhanced by a factor of $\approx 3 \approx 1+\beta^2$ for $\Psi =\pm \pi/2,0$ and were measured to be $T_y=641 \pm 0.2$ mK and $T_x=936 \pm 3$ mK
 respectively,  as expected.

The results show that positive detection of the distinctive $S_{xy}$ shape suffices as a  signature of directionality.   Sensitivity analysis is, however,  useful for more precise measurement of $\Psi$.  Details are found in in appendix,  but noting that the variance errors are similar; and that the $S_{xy}$  have the advantage of excluding a signal from $S_{\text{ff}}^{\text{th}}$; then, a key element is to compare the relative gain of force sensing with $S_{xy}$ ,  compared to PSDs,  by considering the ratio (for $\omega \simeq \omega_x$):

\begin{equation}
\label{RelGain}
\frac{S_{xy}(\omega)}{S_{xx}(\omega)} = \frac{\text{Re}[\chi_x(\omega)\chi_y^*(\omega)]}{|\chi_x(\omega)|^2} \frac{\beta^2 \sin \Psi \cos \Psi}{1+\beta^2 \cos^2 \Psi } 
\end{equation}
We  approximate $\frac{\text{Re}[\chi_x(\omega)\chi_y^*(\omega)]}{|\chi_x(\omega)|^2} \approx \frac{\Gamma_x^2}{4(\omega_x-\omega_y)^2 +\Gamma_x^2} $  near $\omega \sim \omega_x$.    

Hence,  we readily see that optimal visibility of the cross correlation spectrum,  relative to the ordinary PSD,  requires 
$\Gamma_x  \sim  |\omega_x-\omega_y|$.  As the gas damping is very weak,  this means that cooling (such as by active feedback) is essential or
else the cross correlations are strongly suppressed.  A similar result is obtained at $\omega \simeq \omega_y$, by comparing $S_{yy}$ and $S_{xy}$.  
We note that,  in contrast,  standard  1D force sensing is insensitive to cooling rate $\Gamma$ (see  Appendix for further details).

 \begin{figure}[ht]
{\includegraphics[width=3.5in]{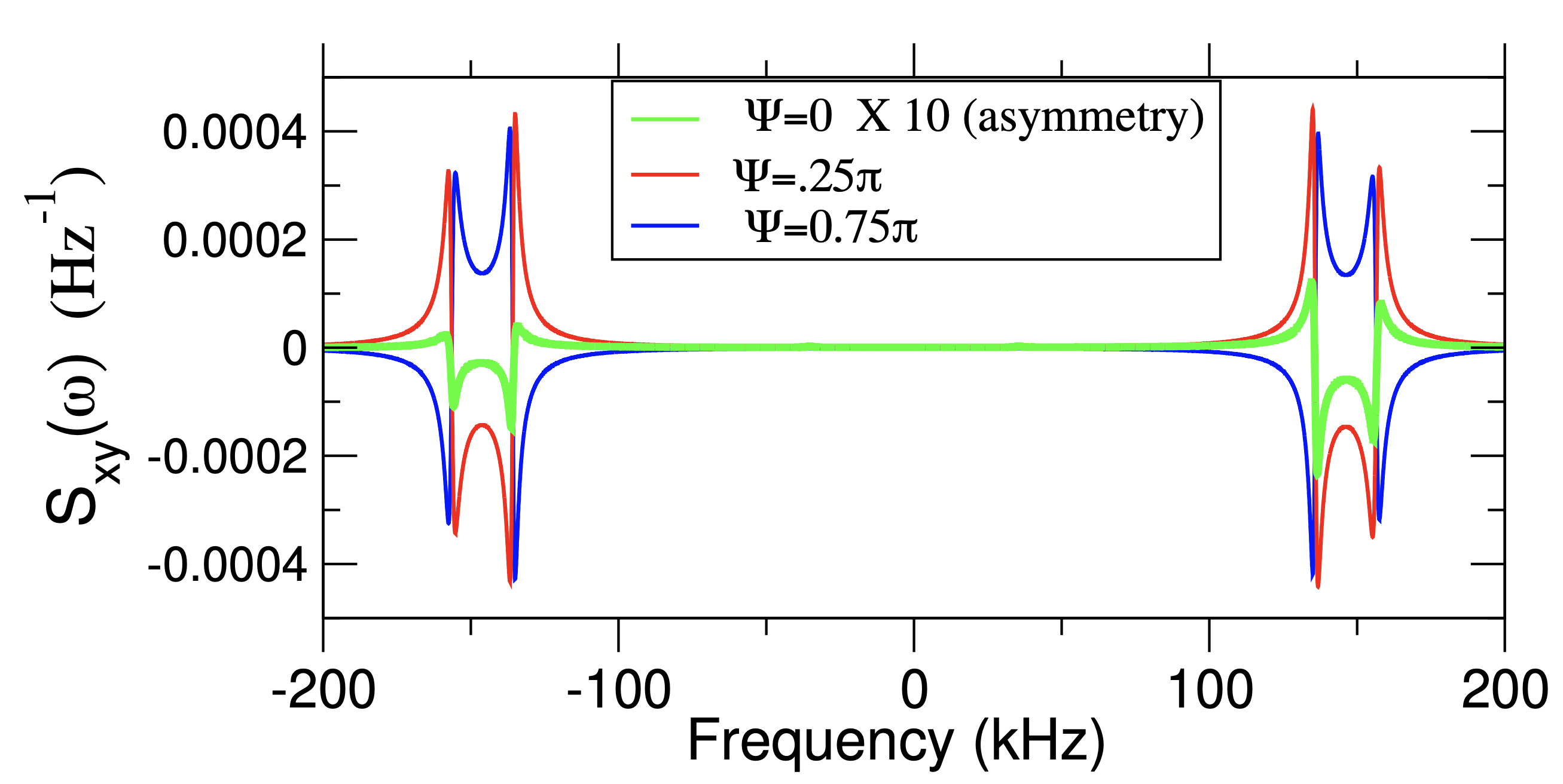}} \caption{Spectral asymmetry even far from quantum regimes: a levitated particle in an optical cavity 
subject to a classical thermal bath with anisotropy $\beta^2=1/4$.  The mechanical modes $x,y$ are equally coupled to the optical mode of the cavity
and have occupancies $n_{x,y}\sim 25$.   For orientations angles $\Psi=0, \pi/2$ the classical bath correlations vanish, exposing the effect of the quantum optical shot noise bath and resulting in a very noticeable asymmetry even when the $x,y$ modes are far from the ground state.
Cavity parameters from the set-up in \cite{Pontin2023}, with gas pressure $P= 3 \times 10^{-6}$\,mbar and $\beta^2=1/4$. }
\label{Fig3}
\end{figure}

\section{quantum vs classical noise baths}

The case of the nanoparticle trapped in a cavity is especially interesting as the particle is subjected to the classical noise bath (from the background gas) simultaneously with  a quantum bath (optical quantum shot noise).

The  experiments presented here are in free-space and in these set-ups,   2D quantum cooling has not yet been achieved.  While future technical advances may allow 2D quantum cooling  without  a cavity and  direct measurement of $S_{xy}(\omega)$,  to date only cavity setups have achieved quantum regimes in the $x-y$ plane,  with the coherent scattering set-up ~\cite{delic2020cooling} that allows very strong light-matter coupling rates $g_x,g_y$.  

Hence one might apply the technique employing 2D quantum cooling in a cavity,  provided split detection offers adequate sensitivity.  However 
this opens up another problem which is that cavity-mediated coupling between modes leads to  hybridisation of the $x-y$ modes and the formation of dark modes \cite{MTTM2020, MTTM2021,Marin2021} and back-action induced rotation of the $x-y$ modes.   2D quantum cooling occurs in a `Goldilocks' zone of  coupling that is not  large enough to form dark modes,  but remains large enough for effective cooling \cite{MTTM2021,Novotny2023}.  To avoid misalignments between the detectors and the $x,y$ modes,  the orientation of the normal modes must be identified, for each different set of experimental parameters. Alternatively,  one may  use a trapping point where the back-action rotation is suppressed \cite{Pontin2023} and the normal modes remain aligned with the tweezer axes.  For the simulations here we consider the nanoparticle in this unperturbed trapping point, investigated experimentally in \cite{Pontin2023}.

A directional component in the Brownian gas collisions is introduced, with $\beta^2=1/4$ and the PSDs and cross-correlations may be obtained analytically using the standard methods Quantum Linear Theory (QLT).   
In Fig.~\ref{Fig3} we plot $S_{xy}(\omega)$  for the cavity parameters in \cite{Pontin2023}  except now pressure $P=3 \times 10^{-6}$\,mbar.
Derivations are in  appendix E where expressions for $S_{xy}(\omega)$ 
plotted in Fig.~\ref{Fig3} are given in Eq.(\ref{Corr2}).  

The $x,y$ modes which are equally coupled to the light field have a phonon occupancy of $n_x, n_y \simeq 25$, thus the ordinary PSDs display little asymmetry.  We note that sideband asymmetry of $S_{xx},S_{yy}$ is an extremely well-studied experimental signature of quantum  cooling since the area of the positive (negative) frequency peaks is proportional to $n_{x,y}$,($n_{x,y}+1$) respectively.

However, the behaviour of the cross-correlations is remarkable:  for most orientation angles $\Psi$ the $S_{xy}$ sidebands are symmetrical; however for $\Psi \sim 0,\pi/2...$ the cross-correlation from the Brownian gas motion is suppresses and the spectra should vanish.
At this point, the weak contribution from the quantum optical shot noise bath is exposed and a striking degree of asymmetry is seen
even if the nanoparticle is quite far from the quantum ground state.
This is in sharp contrast with PSDs where strong asymmetry is only ever present as phonon occupancies $\to 0$.

\section{ Conclusions:}

 We  propose and demonstrate for the first time  that the steady state cross-correlation spectrum  provides a distinctive and unambiguous signature of the presence 
of a directed stochastic force and   its orientation quadrant.  The experiment  exploited a singular advantage of levitated nanoparticles wherein  its centre of mass motion modes have a clear geometric interpretation (motion  along $x,y,z$ axes \cite{MTTM2020,MTTM2021}),  unlike,  say,  two breathing modes of a  membrane.
The method generalises to other planes,  provided the condition  $\Gamma_{i,j} \sim |\omega_i-\omega_j|$ (where $i, j \equiv x,y,z$) can be satisfied.

 In principle,  by measuring relative changes in $S_{xx}$ and $S_{yy}$,  measured independently,  one might infer  information on directionality;
but such changes might be indistinguishable from fluctuations in the underlying experimental heating and cooling rates.  And as the normal PSDs cannot discriminate between orientation quadrants (thus between $\pm \Psi$),  some information is lost.   However,  as measuring $S_{xy}$ implies measurement of 
 $S_{xx}, S_{yy}$,  no choice is required: the cross-correlation complements current methods at no extra cost.  There is  the added advantage that it is not necessary to subtract the unknown uncorrelated component  as $S_{xy}$  isolates $S_{ff}^{dir}$: i.e. for our artificial stochastic noise source, $\beta$ was specified accurately; in contrast,  for uncontrolled noise,  $S_{th}$  adds an uninteresting and uncertain component to the signal (see Appendix).
Sensitivity  may be enhanced by improving the cooling to frequency ratio which is not optimal as $\Gamma_{x,y} / |\omega_x-\omega_y|\sim 1/7 $; and operating at lower pressures of $10^{-6}-10^{-8}$ mbar , near quantum limited regimes.   The method is also  relevant to narrowband directed forces,  provided  $\Gamma_{cool} \sim |\omega_x-\omega_y|$. 

Our study opens the way to detection of different classes of forces of microscopic origin such as tiny gas currents,  small anisotropies in the background noise baths,  possibly due to incoming streams of gas or temperature variations~\cite{Millen2014}.  There is also the prospect of fundamental applications:  the search for a suitable dark matter candidate has not yet been successful and remains one of the major  challenges of physics.   Directionality is expected and significantly increases background rejection,  even for a handful of events \cite{ahlen2010case}.  Searches for dark matter within optomechanics focus on detection of higher energy single recoils \cite{Carney2021a},  or ultralight,  weak  but narrowband sources \cite{Carney2021b}.    However,  weak but broadband directional sources might also be considered \cite{centers2021stochastic}.   This possibility might encompass an ultralight source with a very short coherence time; or a source spanning a band of multiple frequencies.   The natural charge of levitated nanospheres (from 1 to 1000's elementary charges) also opens the possibility of searches of millicharged dark matter \cite{MilliCharges2023} with levitation in ion traps \cite{Millen2015,Bullier2020,Pontin2020}.

{\em Acknowledgements:} The authors would like to acknowledge helpful discussions with Markus~Rademacher, Hayden~Fu, Fiona~Alder  and Marko~Toro\v{s}.  JHI acknowledges an EPSRC DTP studentship. The authors acknowledge funding from the Engineering and Physical Sciences Research Council (EPSRC) Grant No. EP/N031105/1.
\bibliographystyle{unsrt}
\bibliography{2Dbiblio}

\appendix

\section{Sensitivity analysis}

In this  section we present details of  sensitivity analysis for directional force sensing using cross-correlations, relative to the PSD approach.
A standard signal to noise analysis is complicated by the fact that our `signal' is the expectation value of a stochastic force with similar spectral characteristics to the thermal noises that act on the nanoparticle.  Not only do we need to exclude the uninteresting isotropic contribution,  but this also means the distinction between signal and noise is less clear-cut than for, say, detection of a harmonic drive.  Nevertheless we may still apply  techniques widely used  in optomechanics-based force sensing. We present a brief review of standard methods, then apply that to cross-correlations.
Finally we present further details of applications.

\subsection{ Force sensing in levitated optomechanics: 1D analysis and  minimum detectable force}

There is already a considerable body of work on ultra-sensitive force sensing ~\cite{millen2020optomechanics, Geraci2016,hebestreit2018sensing,dark_matter_3, liang2023yoctonewton, hempston2017force,timberlake2019static} with levitated nanoparticles.
A central result encapsulates measurement sensitivity  in an expression for the minimum detectable force in terms of Boltzman's constant, $k$,  temperature $T_x$ of center of mass motion along $x$ and mass of the nanoparticle $m$:
\begin{equation}\label{Fmin}
 F_{min}= \sqrt{ 2 k T_x m \Gamma b}.
\end{equation}
$b$ is the measurement bandwidth.  The expression  sets,  as a hard limit,  that the minimum detectable force signal must exceed the variance of the noise arising from thermal fluctuations  from sources such as for instance, the Brownian noise from background gas collisions.

 In order to justify Eq.\ref{Fmin}, we  consider a 1D damped harmonic oscillator subject to a fluctuating force $F_x$ that is to be measured:

\begin{equation}
m \left[\ddot{x}+\Gamma_x \dot{x}+\omega_x^2 x\right]=F_x(t)
\end{equation}

Analysis of the response of the oscillator to the driving is via the well-known transformation to Fourier space.
If the driving force is deterministic,  one obtains:
\begin{equation}
 x(\omega)= \chi_x(\omega) F_x(\omega)
  \label{ForceX}
\end{equation}
 
where $\chi_x=[m(\omega_x^2-\omega^2+i \omega \Gamma_x)]^{-1}$ is the force susceptibility.

Stochastic forces require analysis via the power spectral densities (PSDs).
Regardless of whether $F_x(t)$ is deterministic or stochastic,  provided steady state is attained,  we can write:
\begin{equation}
  S_{xx}(\omega) =|\chi_x(\omega)|^2 S_{\text{ff}} (\omega)
 \label{Force}
\end{equation}
where  $S_{\text{ff}} (\omega)$ is the PSD of $F_x(t)$.

In order to understand Eq.\ref{Fmin},    we consider the case where $S_{xx}(\omega)$  is driven purely by thermal
baths corresponding to thermal motion at a temperature $T_x$.  The corresponding thermal
force fluctuations are uncorrelated in time $\langle F_x(t)F_x(t')\rangle= S_{\text{ff}} \delta(t-t')$   yielding a
  a white noise  spectrum $S_{\text{ff}}(\omega) \equiv S_{\text{ff}}$.
 The variance of $x$ in steady state is thus $k T_x = m\omega_x^2 \langle x^2\rangle$ and may be obtained
  by integrating over the PSD:

\begin{eqnarray}
\langle x^2\rangle & =& \frac{1}{2\pi} \int_{-\infty}^{+\infty} S_{xx}(\omega) d\omega =  \frac{S_{\text{ff}} }{2\pi} \int_{-\infty}^{+\infty} |\chi_x(\omega)|^2 d\omega \nonumber \\
&= & \frac{S_{\text{ff}} }{2m^2\omega_x^2 \Gamma_x}
\end{eqnarray}
hence

\begin{equation}
 \langle x^2\rangle= \frac{kT_x}{m\omega_x^2}=  \frac{S_{\text{ff}}}{2m^2 \omega_x^2 \Gamma_x}
\end{equation}

from whence we readily obtain  $S_{\text{ff}}= 2 k T_x \Gamma_x m$.
However, the corresponding  force $F$ depends on the frequency bandwidth $b$ of the measurement so
writing $\langle F^2\rangle \equiv  S_{\text{ff}} b$ recovers Eq.\ref{Fmin} for the minimum detectable force.\\
 In  detection of narrowband forces,  the bandwidth $b$ is generally  equated with the inverse integration time,  thus  $b\equiv 1/T$.   It may be reduced to improve the variance error.  Reducing $b$ has for instance allowed a demonstration of  yoctoNewton force sensitivity in levitated nanosystems:  a recent study that took integration times up to $T=36 000$s \cite{liang2023yoctonewton}  achieved sensitivity of tens of yoctonewtons.  It was found that system stability ultimately limited the duration of the experimental integration time and thus the minimum detectable force.

\subsection{Variance of force spectra }

When measuring deterministic forces,  $S_{xx}(\omega)$ represents the variance on the measured displacement $x(\omega)$ and  $ \langle F^2\rangle$ is the variance on the inferred force $F(\omega)$.   However since here we are measuring a purely stochastic force,  we rely on Eq.\ref{Force},  rather than 
Eq.\ref{ForceX},  we are  instead interested in the variance of the measured  power spectra  of force $S_{\text{ff}} (\omega) \equiv S_{xx}(\omega)/|\chi_x(\omega)|^2$.   However we may still compare uncertainties in measurement of force power spectra of narrowband deterministic forces with detection of stochastic white force noise  spectra.\\

\begin{figure*}[ht!]
 {\includegraphics[width=6.6in]{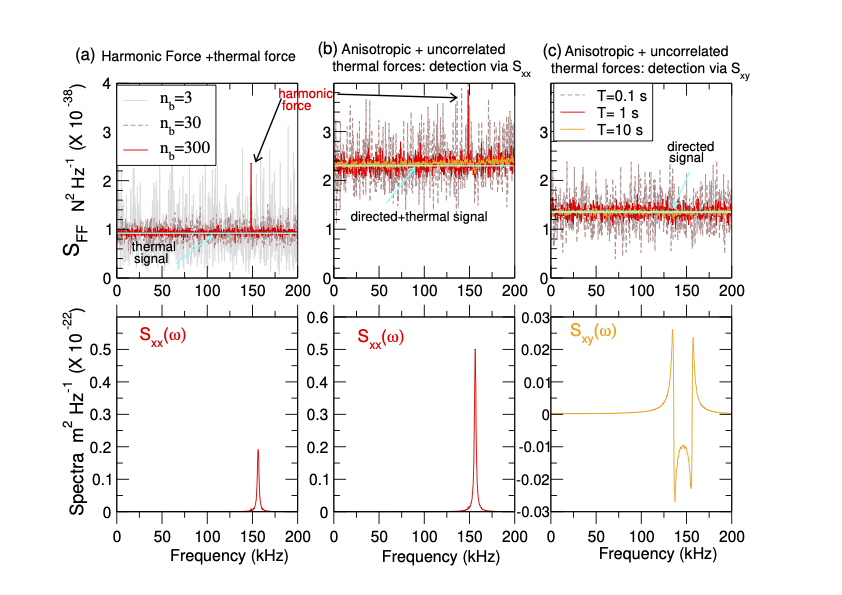}} \caption{ Illustrates optomechanical force sensing. Lower panels show displacement spectra, upper panels show the inferred force spectra. \textbf{(a)} (upper panel) shows force spectra  for a nanoparticle subject to isotropic thermal forces plus a harmonic drive at 149 kHz. As the integration time increases ($T=T_b n_b$, $T_b=3.3$ms), the variance decreases and this sets the minimum detectable force  spectrum.   The harmonic signal shows as a sharp peak.  In contrast, any change in the   (Gaussian white noise) force signal will show as a displacement of the noise floor level.  In both cases the signal must be detectable above the fluctuations with  $\sigma_x \sim  S^{th}_{\text{ff}}/\sqrt{n_b}$,  and this sets the threshold for minimum detectable force power spectrum.  If measuring Gaussian noise, the `signal' of interest corresponds to the $n_b\to \infty$  infinite time integration limit (cyan line).  \textbf{(b)}  (upper panel) shows corresponding force power spectrum
  for a nanoparticle subject to (i) Gaussian noise with an anisotropic component ($\beta^2=3,\Psi=\pi/4$) in addition to the uncorrelated thermal noise (ii) the same harmonic drive along $x$. The standard deviation, for finite $n_b$, is larger than in (a) by a factor $1+\beta^2/2$,  as both noise components contribute.
The white noise floor `signal'  (cyan) is now the sum of the directed plus isotropic components.  As it does not distinguish between them, measuring the anisotropic component has an added uncertainty. The harmonic force signal is once again included to show that, in contrast, for narrowband force sensing, the distinction between signal and the noise floor can be  clear-cut, as long as the variance is sufficiently low  \textbf{(c)}  (upper panel) shows the advantage of using the cross correlations, relative to the PSDs. For a nanoparticle subject to the same combined anisotropic plus uncorrelated thermal forces as in (b), detection via cross-correlation isolates the directed anisotropic component of the noise, eliminating a source of uncertainty in the signal.  The noise variance includes both noise contributions so is, as expected, similar for (b) and (c).
  Pressure is $P=1\times 10^{-3}$ mbar. }
\label{FigS1}
\end{figure*}

In Fig.\ref{FigS1}  we illustrate and contrast detection of  narrowband forces with detection of  stochastic white-noise forces
for different $T$. 
 We employed stochastic solutions of the time dependent equations of motion for $x,y,z$, in the presence of external forces, and  evolved the dynamics until steady state was attained.  The lower panels show displacement spectra (for both PSDs and cross-correlations).  The upper panels show the corresponding inferred force spectra.
 
We consider first a force with combined  harmonic plus white-noise thermal components: 
 \begin{equation} \label{Fharm}
 F_x(t)=F_x^{harm}(t)+ f_{\text{th},x}(t)
 \end{equation}
and where $F_x^{harm}= F_0 \cos \omega_h t$.  As the system is linear, the corresponding spectra are additive
 $S_{\text{ff}} (\omega)= S^{th}_{\text{ff}}+ S^{harm}_{\text{ff}} (\omega)$.
 Our measured force spectrum is at finite integration time $T$, hence: 
\begin{equation}
S_{\text{ff}}(\omega,T) \equiv \langle |x(\omega,T)|^2\rangle / |\chi_x(\omega)|^2 \simeq S^{th}_{\text{ff}}+ S^{harm}_{\text{ff}} (\omega)
\end{equation}

$S_{\text{ff}}(\omega,T)$ fluctuates significantly for finite $T$.  The associated variance error might in principle be arbitrarily mitigated by increasing $T$ and averaging neighbouring frequency bins.
 In practice  the measured time traces are split into  $n_b$ independent integration blocks of length $T_b$, thus $T=n_bT_b$. In that case $T\to \infty $ is approached as $n_b \to \infty$ and the frequency bin width is $1/T_b$: in all calculations below we fix $T_b$ and vary $n_b$.

Our measured spectrum is a discretised  force spectrum calculated by averaging over an ensemble of $n_b$
measured or simulated data points:
\begin{equation}
S_{\text{ff}}(\omega_j,T)= \frac{1}{T_bn_b}  \sum_{l=1}^{n_b}     x_l^*(\omega_j) x_l(\omega_j)/|\chi_x(\omega_j)|^2
\end{equation}
which fluctuates, with a variance $Var[S_{\text{ff}}(\omega,T)] $,  about the expectation value given by
 $S_{\text{ff}}(\omega_j,T\to \infty) \equiv \langle S_{\text{ff}}\rangle$ , which is frequency independent for a white noise force.  The corresponding variance of the force measured with integration time $T$ is:
\begin{equation}
Var[S_{\text{ff}}(\omega,T)] \equiv \langle \  \left\langle S^{2}_{\text{ff}}(\omega,T)\rangle- \langle S_{\text{ff}}\rangle^2  \ \right\rangle
\end{equation}
where the average is over the usable frequency bandwidth (in principle infinite,  in practice limited by measurement noise to a  finite range $\Delta f$,  as discussed below).

  In Fig.\ref{FigS1} (a) we  compare the PSD ($S_{xx}(\omega)$) (lower panel) with the corresponding estimated force spectrum $ S_{xx}(\omega)/|\chi_x(\omega)|^2$ (upper panel).  It can be seen that $F_x^{harm}$  results in a sharp feature $\propto F^2_0 \delta(\omega-\omega_h)$.   The thermal component, on the other hand,  appears in the force spectrum as a  noise `floor' at $S^{th}_{\text{ff}} \simeq 2k m T_{x,y} \Gamma_{x,y}$,   which is  flat but with a strongly fluctuating background- a variance due to thermal fluctuations that is reduced as $n_b\to \infty$.

Fig.\ref{FigS1} (a)  illustrates  the minimum detectable force  spectrum and the role of $b$ and integration time.
 The minimum detectable force must exceed the variance error.  The delta peak `signal' due to the harmonic drive must exceed this noise to become detectable: it is clearly the case for $n_b=30$ (red line) but not for $n_b=3$ (grey line).

Since in the present work we  are interested in sensing a component of the Gaussian noises, we are effectively detecting a change in the noise floor level.   The `signal' of interest  is actually the flat  $S_{\text{ff}}\equiv S^{th}_{\text{ff}}$    thus the limit of long integration time, $T \to \infty $.  This change in the floor level too must exceed the error $\sigma_x$.

In summary: for the simple  damped harmonic oscillator in Fig.\ref{FigS1} (a),  for detection of thermal noises, the `signal' corresponds to  $ S^{th}_{\text{ff}}= 2k m T_x \Gamma_x$ and  $\sigma^2_x[S_{xx}(\omega)/|\chi_x(\omega)|^2] = Var[S_{\text{ff}}(\omega,T)] \simeq   [S^{th}_{\text{ff}}]^2/n_b$,  so our  error improves, as  $\sigma_x \propto  1/\sqrt{n_b}$,  and as the integration time is increased.  

 \section{Sensing anisotropic noise}

 We now consider the case where the driving force is purely stochastic but one component is able to induce correlations between two degrees of freedom $x,y,z$ of the centre of mass motion.  We consider $x$ and $y$.
For the case of centre of mass motion,  in our work we note that this may be straightforwardly related to directionality of the stochastic bath.
 \subsection{2D analysis with PSDs}

We consider that our system is exposed to external forces comprising a harmonic driving force,  a thermal, randomly oriented,  component and now,  an additional  stochastic,  directed
Gaussian white noise component:

 \begin{eqnarray} \label{Fdir}
 F_x &=& f^{dir,x}(t)+ f_{\text{th},x}(t)=f^{dir}(t) \cos \Psi + f_{\text{th},x}(t) + F_x^{harm}\nonumber \\
 F_y & = & f^{dir,y}(t)+ f_{\text{th},y}(t)=f^{dir}(t) \sin \Psi +  f_{\text{th},y}(t).
 \end{eqnarray}

We attempt to estimate the external forces from the displacement PSDs corresponding to integration time $T$:
\begin{eqnarray}
 \frac{S_{xx}(\omega,T)} {|\chi_x(\omega)|^2}  &\simeq &  S^{dir}_{\text{ff}} \cos^2 \Psi  + S^{th}_{_{\text{ff}},x} + S^{harm}_{_{\text{ff}},x}\nonumber  \\
  \frac{S_{yy}(\omega,T)} {|\chi_y(\omega)|^2}  &\simeq&  S^{dir}_{\text{ff}} \sin^2 \Psi + S^{th}_{_{\text{ff}},y} \nonumber \\
    \frac{S_{xy}(\omega,T)} {\text{Re}[\chi_y^*(\omega)\chi_x(\omega)] } &\simeq &  S^{dir}_{\text{ff}} \sin \Psi \cos \Psi
    \label{PSDxy}
 \end{eqnarray}
where we introduce  the notation $S_{xx}(\omega,T) \equiv  \langle |x(\omega,T)|^2\rangle$ to denote the finite time ensemble averaging
in the experimental spectra. Although  we assume that $T$ (or $n_b$) is large enough that we can  provide a reasonable estimate of the force spectrum, it is still a fluctuating signal, with a non-zero variance.

The experiments here employed a controlled,  artificial directed stochastic source 
and as the isotropic and anisotropic noises are both white,   we can model the effect of the applied stochastic forces  by specifying the ratio:
 \begin{equation}\
\beta^2 = S_{\text{ff}}^{dir}/S^{\text{th}}_{\text{ff}}.
\end{equation}
One would not expect to know $\beta$ a-priori in a realistic sensing scenario,  thus it should in general be treated as an unknown parameter: one may not be able to  `turn off' the directed component so as to calibrate the experiment for 
$S^{th}_{\text{ff},x},S^{th}_{_{\text{ff}},y}$.  The effect of adding the anisotropic/directed component in the PSD is to shift the noise floor; thus detecting the extra force $ S^{dir}_{\text{ff}} \cos^2 \Psi $ requires one  to estimate that shift relative to $ S^{th}_{\text{ff}}$,  adding an uncertainty to the signal.

This shift  is illustrated in  Fig.\ref{FigS1}(b) (upper panel) , where we once again plot the PSD and the corresponding total $x$ force spectrum, for the case where  $\beta^2=3$.  
For contrast,  a narrowband signal such as a harmonic drive is also shown: $S^{harm}_{ff,x}$  appears as a delta-peak $\delta(\omega-\omega_h)$,  at the drive frequency.   Here the distinction between the near delta-spike harmonic force signal and the white noise floor is clear-cut and a simple shift in the noise floor level does not preclude measurement of the amplitude of the harmonic signal.

The variance for the same given integration time $T$ has to allow for the additional anisotropic noise bath. Hence, $Var[S_{xx}(\omega,T)/|\chi_x(\omega)|^2 ] \simeq  [(1+\beta^2\cos^2 \Psi)  S^{\text{th}}_{\text{ff}}]^2/n_b$ while 
$Var[S_{yy}(\omega,T)/|\chi_y(\omega)|^2] \simeq  [(1+\beta^2\sin^2 \Psi)  S^{\text{th}}_{\text{ff}}]^2/n_b$. 

In order to  measure $\Psi$   we can subtract the thermal contributions:
\begin{eqnarray}
 \frac{S_{xx}(\omega,T)} {|\chi_x(\omega)|^2} -   S^{th}_{_{\text{ff}},x}  &\approx &  S^{dir}_{\text{ff}} \cos^2 \Psi \equiv \Delta S^x  \nonumber  \\
    \frac{S_{yy}(\omega,T)} {|\chi_y(\omega)|^2} - S^{th}_{_{\text{ff}},y}  &\approx & S^{dir}_{\text{ff}} \sin^2 \Psi  \equiv  \Delta S^y \\
    \label{force1}
 \end{eqnarray}
then,  even   if the magnitude of the anisotropic component $ f^{dir}(t)$ is completely unknown,  we  combine the shift in $x$ with the shift in $y$ to estimate $ \Psi$ from the ratio:
\begin{equation}
\label{ratio}
\Delta S^y/\Delta S^x \simeq  \tan^2 \Psi
\end{equation}
Eq.\ref{ratio} exposes the fact that any sort of measurement using the PSDs is insensitive to  $\pm \Psi$.
In addition,  Eq.\ref{force1} exposes a further  source of uncertainty as we must measure two thermal components to extract $\Psi$: 
 the ratio of $S^{th}_{ff,x}$ and $S^{th}_{ff,y}$ is generally not accurately known and must be carefully calibrated.  The thermal components contain not only the effects of the isotropic gas collisions,  but also $x,y$ channel-dependent electronic noises, arising from the feedback cooling for instance,  that can shift not only the magnitude but the ratio of the $x,y$ components.  These can vary significantly over different experimental runs.  Calibration may be complicated by the fact that one may not (as in this work)
controllably turn the added anisotropy on and off for comparison.  Pressure gauges may have $10-30\%$ uncertainties which would translate into a corresponding scaling of the thermal component, undistinguishable
from adding $f^{dir}$.  In general  nanoparticles are not perfectly spherical, thus $\Gamma_x \neq \Gamma_y$ \cite{Rademacher2022} 
and the response is not necessarily identical for $x,y$ if a different nanoparticle is loaded into the trap.\\

{\em In summary:}  PSDs are in effect `blind' to whether the additional component is directional anisotropy or whether it is a channel dependent additional noise.  The requirement to characterise and subtract the uninteresting uncorrelated thermal component,  for each experimental run,  can be challenging and adds a potentially significant uncertain component to the measured signal.

  \subsection{Noise anisotropy: cross-correlation spectra}

  \subsubsection{Calibration-free  detection of a directed component}
  In Fig.\ref{FigS1}(c) (lower panel) we show the cross-correlation spectra and inferred force using Eq.\ref{PSDxy}.
  The figure shows the characteristic shape that indicates that a directional/anisotropic noise is present.
 In the absence of a directed noise,  $\langle S_{xy}(\omega)\rangle=0$.
   As shown in this work detection and analysis of $S_{xy}$ augments the PSD-measurement as it can evidently discriminate between isotropic and directed noise.  As the signature shape `flips' between $\pm \Psi$,  it offers also calibration free
   discrimination between quadrants.

   A detailed sensitivity analysis is not essential for this key result of our work.

    The simple observation of a detectable $S_{xy}(\omega,T)$  trace suffices to unambiguously signal the presence of a directed component.  A properly set-up apparatus
   is of course assumed: misalignments between the detection $x,y$ axis and the normal modes of the motion
   can introduce spurious cross-correlations; fortunately free-space experiments do not introduce the back-action induced rotation of normal modes that are a feature of cavity set-ups \cite{Pontin2023}.  But the signature shape is insensitive to re-loading of a new nanoparticle,  fluctuations in optical power, in feedback voltage or background gas density.

    \subsubsection{Detection of force orientation with calibration}

 In Fig.\ref{FigS1}(c) (upper panel) we extract the directed force spectrum from the cross-correlations: 

 \begin{equation}
    \frac{S_{xy}(\omega,T)} {\text{Re}[\chi_y^*(\omega)\chi_x(\omega)]}  \simeq  S^{dir}_{\text{ff}} \sin \Psi \cos \Psi
    \label{PSxy}
 \end{equation}
The anisotropic/directed component is now isolated,  without the need to estimate $\beta$.
Comparison between the upper panels in Fig.\ref{FigS1}(b) and (c) 
illustrates that the variance `noise' on the force is similar relative to that obtained from PSDs.  This seems intuitive
 and reasonable: the cross-correlations are constructed from precisely the same basic components as the PSDs,
 ie the measured time series $x(t), y(t)$.  
 
 No harmonic force signal is seen as $F_x^{harm}$ is along $x$.   A harmonic force not perfectly aligned with $x$ or $y$ would  also yield a sharp peak in the cross-correlation,  but in this work we focus on broad-band force detection so do not include a 
deterministic drive in  Fig.\ref{FigS1}(c).

 The variance of the cross-correlation may be written  $Var[|\langle x^\dagger y\rangle|]=\sqrt{Var[S_{xx}(\omega,T)]Var[S_{yy}(\omega,T)]}$   \cite{Bendat}.   Squaring both sides and dividing both sides  by $ |\chi_y(\omega)|^2 |\chi_x(\omega)|^2$, we obtain:
  \begin{eqnarray}
Var[ S_{\text{ff}} (\omega,T)] & \simeq &  Var[\frac{|\langle x^\dagger y\rangle|}{|\chi_y(\omega)| |\chi_x(\omega)|}]\nonumber \\
 &\simeq& {(1+\beta^2\cos^2 \Psi)  (1+\beta^2\sin^2 \Psi)}(S^{\text{th}}_{\text{ff}})^2/n_b.
\label{varSxy}
 \end{eqnarray}
 
for the variance of the force spectrum extracted from the cross-correlation.
Note that here we measure and investigate $S_{xy}= \frac{1}{2}[  \langle x^\dagger y\rangle + \langle x y^\dagger\rangle]$  
which corrresponds to $\text{Re}\langle x^\dagger y\rangle $ but we can assume
 the  real and imaginary components of $\langle x^\dagger y\rangle$ have similar statistical behavior.  

 
{\em  In summary:} Eq.\ref{varSxy}  shows that the variance of the force spectrum inferred from the cross-correlation contains both the directed and isotropic contributions so  is of similar magnitude to that
 inferred from the PSDs.  The mean or expectation value (the noise floor, for $T\to \infty$), in contrast,  isolates the directed contribution.
 We must now consider the effect of the imprecision noise: there is no perfect measurement of
 the PSDs or cross-correlation; they are both also affected by measurement/imprecision noises,
 but we show that the disparity in gain implies the effect is not equal.

 \subsubsection{Imprecision noise}

In a realistic experiment, the motion is detected exploiting interferometric techniques which will have an associated imprecision noise. Additionally, electronic noise can potentially be added through the transduction chain when going from optical to electronic signal. In principle one can write the measured electronic signal $v(t)$ as given by $v(t)=g_{el} (g_{opt} x(t)+s_{shot})+s_{el}$. Here, $g_{opt}$ is the transduction coefficient from displacement to optical intensity, $g_{el}$ is the photodetection gain while $s_{shot}$ and $s_{el}$ are stochastic variables associated with optical shot noise and electronic noise respectively. Upon calibration, the signal $v(t)$ provides a measurement of the particle motion. 

A typical and simple way to describe this situation is to consider that a measurement of the particle position will have an associated imprecision noise, i.e., $x_m(t)=x(t)+s_{imp}$. We thus write for the measured $x_m$ position PSD
\begin{equation}
 S^{m}_{xx}(\omega) =  S_{xx}(\omega)+ S_{imp}
\end{equation}
where $S_{imp}$ is the PSD of the imprecision noise that for simplicity we approximate to be white noise.  
This is certainly the case for the shot noise contribution. In general, other contributions to $s_{imp}$ might have different color but at least over the interesting frequency range, are approximately flat.

Below we do not include the $T$ and $S^{m}_{xx}(\omega)\equiv S^{m}_{xx}(\omega,T)$ as a finite $T$ is implied in any measurement.



As the imprecision noise is unfiltered by the nanoparticle motion,  it is purely additive. We then obtain for the measured force PSD:
\begin{eqnarray}
  \frac{S^{m}_{xx}(\omega)} { |\chi_x(\omega)|^2}   \equiv   S^{m}_{\text{ff}}   =   S^{dir}_{\text{ff}}\cos^2 \Psi + S^{th}_{_{\text{ff}},x}+ \frac{S_{imp}}  { |\chi_x(\omega)|^2}
 \label{Forcexx}
\end{eqnarray}

Thus even if the system is only driven by a white noise force (i.e., $F_{harm}=0$) and the imprecision noise has a flat spectrum, the measured force spectra now has a strong frequency dependence because of the imprecision noise. But since $|\chi_x(\omega)|^2$ has a sharp maximum for $\omega=\omega_x$ the imprecision contribution $\frac{S_{imp}} {|\chi_x(\omega)|^2}$ is suppressed near $\omega \approx \omega_x$ and also for high gain.

Force sensitivity is thus maximal on-resonance.

The imprecision contribution introduces an effective upper limit on the measurement bandwidth: while for perfectly flat white noise   force spectrum,  we may in principle use the full frequency range to converge to $\langle S_{\text{ff}} \rangle$, in practice the presence of coloured or imprecision noises mean that sensing is restricted to a narrow frequency band $\Delta f$  where $S_{\text{ff}} >\frac{S_{imp}} {|\chi_x(\omega)|^2}$, near resonant frequencies. With this further averaging, over $\Delta f$,  we  may then reduce the variance of $\langle S^{th}_{\text{ff}}\rangle$ as  $1/\Delta f T$.   On the other hand, if $\Delta f$ is too narrow, or even absent,  if $\frac{S_{imp}} {|\chi_x(\omega)|^2} > S_{\text{ff}}$ over the entire frequency range, detection becomes impractical.

Measurement noise clearly affects also force measurement via cross-correlations.  We may  estimate the measured force spectrum,  $S^{m,xy}_{\text{ff}}$, from  measured  cross-correlations,  $S^{m}_{xy}(\omega)$,  as follows:

\begin{eqnarray}
   \frac{S^{m}_{xy}}  {\text{Re}[\chi^*_x(\omega)\chi_y(\omega)]} & = & S^{m,xy}_{\text{ff}}  \\
   &  =  & S^{dir}_{\text{ff}}\cos \Psi \sin \Psi + \frac{S_{imp}}  {\text{Re}[\chi^*_x(\omega)\chi_y(\omega)]}.\nonumber 
 \label{Forcexy}
\end{eqnarray}

Our Eq.(\ref{Forcexx}) and (\ref{Forcexy}) are analogous but expose clearly the relative advantages: Eq.\ref{Forcexy} does not extract the uncorrelated component, thus eliminating that uncertainty in the signal.
On the other hand,  as the effective gain $\text{Re}[\chi^*_x(\omega)\chi_y(\omega)]$ can be significantly lower,  cross-correlations may become undetectable, while the PSDs have the relative advantage of a larger gain $|\chi_{x,y}(\omega)|^2$.  To mitigate this disadvantage,  one must either minimise imprecision noises,  or improve the relative gain function as explained in Eq.(5) and (6) of the main text.

\section{Relative sensitivity of PSD and cross-correlation spectra}

Our measured signal for force detection is:
\begin{equation}
  S^{m}_{xy} (\omega) \simeq  \text{Re}[\chi^*_x(\omega)\chi_y(\omega)]S^{th}_{\text{ff}}\beta^2 \cos \Psi \sin \Psi
\end{equation}
with cross-correlation, and
\begin{equation}
  S^{m}_{xx}(\omega) \simeq   |\chi_x(\omega)|^2 S^{th}_{\text{ff}}(1+\beta^2 \cos^2 \Psi)
\end{equation}
for the $x$ PSD.  The noise standard deviation for both cases is of the same order $\sim \sqrt{(1+\beta^2\cos^2 \Psi)  (1+\beta^2\sin^2 \Psi)} S^{th}_{\text{ff}}/\sqrt{n_b}$  and may be similarly mitigated by longer integration times (larger $n_b$) in both cases.

The  principal disadvantage  for the cross-correlation is that its signal can be associated with a relatively weaker gain function unless   $\Gamma \sim |\omega_x-\omega_y|$,  in which case the gain functions can become of the same order of magnitude.  This follows from  consideration of the ratio:
\begin{equation}
\label{RelGain}
\frac{S_{xy}(\omega)}{S_{xx}(\omega)} = \frac{\text{Re}[\chi_x(\omega)\chi_y^*(\omega)]}{|\chi_x(\omega)|^2} \frac{\beta^2 \sin \Psi \cos \Psi}{1+\beta^2 \cos^2 \Psi }. 
\end{equation}
If we consider this ratio near resonance,   we can approximate $\frac{\text{Re}[\chi_x(\omega)\chi_y^*(\omega)]}{|\chi_x(\omega)|^2} \approx \frac{\Gamma_x^2}{4(\omega_x-\omega_y)^2 +\Gamma_x^2} $  near $\omega \sim \omega_x$,  provided  $\omega_x,\omega_y \gg |\omega_x-\omega_y| $.   
A similar result is obtained for $\omega \sim \omega_y$.

{\em In summary:}  force detection with cross-correlations provides a calibration-free signature of the presence of an external directed force and its orientation quadrant.  As the cross-correlation spectra is constructed from the same components (measured $x(t),y(t)$) as the PSDs, it might also be viewed as a means to augment PSD-based detection without additional effort.  While both methods are equally affected by noise, the cross-correlation detection signal (as opposed to the noise error)  isolates the directed force signal from its uncorrelated signal and is sensitive to the sign of $\Psi$.

\section{Cooling and sensitivity}

The well-known optomechanics expression Eq.\ref{Fmin} for the minimum detectable force in 1D, $F_{min}= \sqrt{2 k T m \Gamma b}$  exposes the interesting result that (in principle) cooling of the nanoparticle motion does not improve sensitivity.  Even if $T= 300$K,   exquisite enhancements in sensitivity result purely from operating at high vacuum (down to pressures $P\sim 10^{-9}$ mbar and below),  so if the heating is determined by background gas $\Gamma\to 0$.

It is normal to apply some feedback cooling to stabilise the motion in each mode ($x,y$) as the pressure is reduced,  thus 
in general there is a mode dependent damping $\Gamma^{x,y}_{eff}= \Gamma+ \Gamma^{x,y}_{fb}$.  For strong cooling (such as is the case in Fig.(2) of the main text),  $\Gamma^{x,y}_{fb} \gg  \Gamma$ where $\Gamma$ is determined by the background gas collisions. The superscripts arise because feedback cooling rates are applied independently to each channel and thus may vary.  To be more precise,  $\Gamma \to \Gamma^{x,y}$ as most nanoparticles are not perfectly spherical thus even if the loaded particle is carefully selected,  in general $\Gamma_x \neq \Gamma_y$.

In that case, the corresponding temperatures:
\begin{equation}
 T^{x,y}_{eff}= \frac{\Gamma^{x,y}}{\Gamma^{x,y}_{eff} }T_{x,y}.
\end{equation}

The minimum force expression is then
\begin{equation}\label{FminEff}
 F^{x,y}_{min}= \sqrt{2 k T^{x,y}_{eff} m \Gamma^{x,y}_{eff} b}.
\end{equation}
from the above, we see that cooling keeps the $T_{eff}  \Gamma^{x,y}_{eff}$ product invariant so the sensitivity is insensitive to cooling.

The scenario with cross-correlations is different.  The constraint $\Gamma_{eff}^{x,y} \sim |\omega_x-\omega_y|$ means that cooling is in general important for measurable cross-correlations.   Minimising the isotropic thermal noise is also important as that adds to the variance of the measurement, as shown above.


\section{Quantum dynamics}

For a standard optomechanical system comprising an arbtrary  2D mechanical oscillator in an optical cavity, 
 we can give the solutions of the quantum Langevin equations
for the quantum mechanical displacement operators,  in Fourier space,  in the form:
\begin{eqnarray}
\hat{x}(\omega) & = M_{x}^{-1}\left[\sqrt{\Gamma}\tilde{Q}_x^{\text{therm}}+i\sqrt{\kappa}g_{x}\mu_{x}\tilde{A}^{\textrm{in}}\right ] \nonumber \\
\hat{y}(\omega) & = M_{y}^{-1}\left[\sqrt{\Gamma}\tilde{Q}_y^{\text{therm}}+i\sqrt{\kappa}g_{y}\mu_{y}\tilde{A}^{\textrm{in}}\right].
\label{1DDisplace1}
\end{eqnarray}
On the right hand side we have operators for thermal noises (classical in general) and for the quantum optical shot noise.

The thermal noises (and $j\equiv x,y$)  take the form
\begin{equation}
\tilde{Q}_{j}^{\text{therm}}(\omega)=\chi(\omega,\omega_{j},\Gamma)\hat{b}_{j}^{\text{in}}(\omega)+\chi^{*}(-\omega,\omega_{j},\Gamma)\hat{b}_{j}^{\text{in}}{}^{\dagger}(\omega),
\end{equation}

For the normal levitated optomechanics scenario,  the thermal noises are dominated by collisions with  surrounding gas molecules at ambient 
temperature so the corresponding bath is highly occupied, i.e.  
$\langle [\hat{b}_x^{\text{therm}}]^\dagger \hat{b}_x^{\text{therm}}\rangle= \bar{n}_x\delta(t-t')$ and where $\bar{n}_x=kT/(\hbar \omega_x)\gg 1$ where $\bar{n}_x$ is the thermal occupancy of the $x$ mode.  Similarly for $y$.

The associated Brownian motion is isotropic, so one assumes the baths are uncorrelated:
\begin{equation}
  \langle  [\hat{b}_x^{\text{therm}}]^\dagger \hat{b}_y^{\text{therm}}\rangle =\langle  [\hat{b}_y^{\text{therm}}]^\dagger \hat{b}_x^{\text{therm}}\rangle  =0
\label{Uncorr}
\end{equation}

In contrast, the quantum optical shot noise:
\begin{equation}
\tilde{A}^{\textrm{in}}(\omega)=\chi(\omega,-\Delta,\kappa) \hat{a}_{\text{in}}(\omega)+\chi^{*}(-\omega,-\Delta,\kappa)\hat{a}_{\text{in}}^{\dagger}(\omega).
\end{equation}
corresponds to a zero temperature quantum bath and $\langle \hat{a}_{in}^\dagger \hat{a}_{in}\rangle= 0$ while
$\langle \hat{a}_{in} \hat{a}_{in}^\dagger\rangle= \delta(t-t')$.

{\em Optical and mechanical susceptibilities:} The $\mu_{j}(\omega)$ are mechanical susceptibilities, while $\eta_c$ is the optical susceptibility. We have the usual mechanical susceptibility $\mu_j(\omega)=\chi(\omega,\omega_j)-\chi^*(-\omega,\omega_j)$ and optical susceptibility  $\eta_c(\omega)=\chi(\omega,-\Delta)-\chi^*(-\omega,-\Delta)$, where, e.g., $\chi(\omega,\omega_x)=[-i(\omega-\omega_x)+\frac{\Gamma}{2}]^{-1}$ and $\chi(\omega,\Delta)=[-i(\omega-\Delta)+\frac{\kappa}{2}]^{-1}$.

Additionally,  in Eq.\ref{1DDisplace1}  we have the normalization factor (for $j\equiv x,y$)
$M_{j}(\omega)=1+g_{j}^{2}\mu_{j}(\omega)\eta_c(\omega)$.  This important term encompasses the cooling effect of the cavity optical mode
and is responsible for reducing the effective temperature of the mechanical mode.

We now lift the assumption that the thermal baths are uncorrelated by introducing an additional component of gas moving along a definite direction, hence we replace $ \Gamma\hat{b}_{j}^{\text{in}}(\omega)\to  \Gamma  \hat{b}_{j}^{\text{in}}(\omega)+ \Gamma_{j,corr} \hat{b}_{j}^{\text{corr}}(\omega)$. The additional component of correlated noises no longer obeys Eq.\ref{Uncorr} but rather, $ \langle  [\hat{b}_x^{\text{corr}}]^\dagger(t) \hat{b}_y^{\text{corr}}(t')\rangle  \propto \delta(t-t')$.

We model the correlated collisions by a physically intuitive model: a force $F(t)$, at $\Psi=45^\circ$  to $x$ implies that $ f_x(t)=f_y(t)$ at arbitrary times; conversely, at $\Psi=135^\circ$  to $x$
implies that $ f_x(t)=-f_y(t)$ at arbitrary times etc. We consider a broad spectrum force in the $\omega \approx \omega_x,\omega_y$ range, and for simplicity take white noise, so expect $ \langle  f_x (t) f_y(t')\rangle  \propto \sin \Psi \cos \Psi \delta(t-t')$.

For simplicity we consider our directed component to be of the same species as the main Brownian gas collisions but represents a modest fraction $\beta^2 <1 $ of the gas (it is no problem to relax this assumption).  Hence we take $\sqrt{ \Gamma_{x}} \sqrt{\Gamma_{y}} =\Gamma \beta^2  \sin \Psi \cos \Psi$.

We note that the quantum optical shot noise is common to both mechanical modes thus it will introduce correlations between them.  In this case there is no directionality involved, however.  

From the solution of the linearised quantum Langevin equations  (see also \cite{MTTM2020}) we readily obtain:

\begin{equation}
  S_{xy} (\omega) =S_{QB}(\omega)+ S_{Class}(\omega) 
\label{Corr2}
\end{equation}

where the first term is the contribution from the quantum bath of the optical shot noise:
\begin{eqnarray}
  S_{QB}(\omega)&=& \kappa g_xg_y |\chi_c(\omega)|^2 \mathcal{M}_{xy}(\omega)   \nonumber \\
  \end{eqnarray}
  
  while the second term is the contribution from the classical,  thermal Brownian bath:
  \begin{eqnarray}
  S_{Class}(\omega)&=&   \Gamma \frac{\beta^2}{2} \sin \Psi \cos \Psi [ (\bar{n}_x+1+ \bar{n}_y+1)\mathcal{M}_{xy}(\omega) \nonumber \\
   & + & (\bar{n}_x+ \bar{n}_y)\mathcal{M}_{xy}(-\omega) ]
   \label{QLTCorr}
\end{eqnarray}

and where 
\begin{equation}
 \mathcal{M}_{xy}(\omega)=\left[ \frac{\mu_x(\omega)\mu_y^*(\omega)}{M_x(\omega)M_y^*(\omega)} + cc\right]
\end{equation}
is real.  We note that both terms are modulated equally by the frequency dependent functions $\mathcal{M}_{xy}(\omega)$.  However, the quantum bath has only component  and is missing the $M_{xy}(-\omega)$ component.

The classical term will vanish if $\sin 2\Psi=0$.  For these angles,  $S_{xy} (\omega) =S_{QB}(\omega)$ the cross-correlation spectra
are dominated by the quantum noise,  even if the particle is not near its quantum ground state. 

One can estimate the relative strengths of the classical and quantum contributions; for $-\Delta \sim \omega_x,\omega_y$, $|\chi_c(\omega)|^2\sim 4/\kappa^2$. Hence

\begin{equation}
 S_{QB}(\omega): S_{f_x f_y}(\omega) \sim \frac{4g_xg_y}{\kappa}:\Gamma \beta^2\sin \Psi \cos \Psi n_B
\end{equation}
where $ n_B= (\bar{n}_x+ \bar{n}_y)/2$. Thus defining a quantum ``cross-cooperativity'' $C_{xy}$ in close analogy to the usual quantum cooperativity:
\begin{equation}
 C_{xy}= \frac{4g_xg_y}{\kappa\Gamma \bar{n}}
\end{equation}
we see that the classical bath will dominate the correlating effects of the quantum shot noise bath if
\begin{equation}
 C_{xy} \lesssim \beta^2 \sin \Psi \cos \Psi.
\end{equation}

\subsection{Heterodyne detection}

\begin{figure}[ht]
{\includegraphics[width=3.3in]{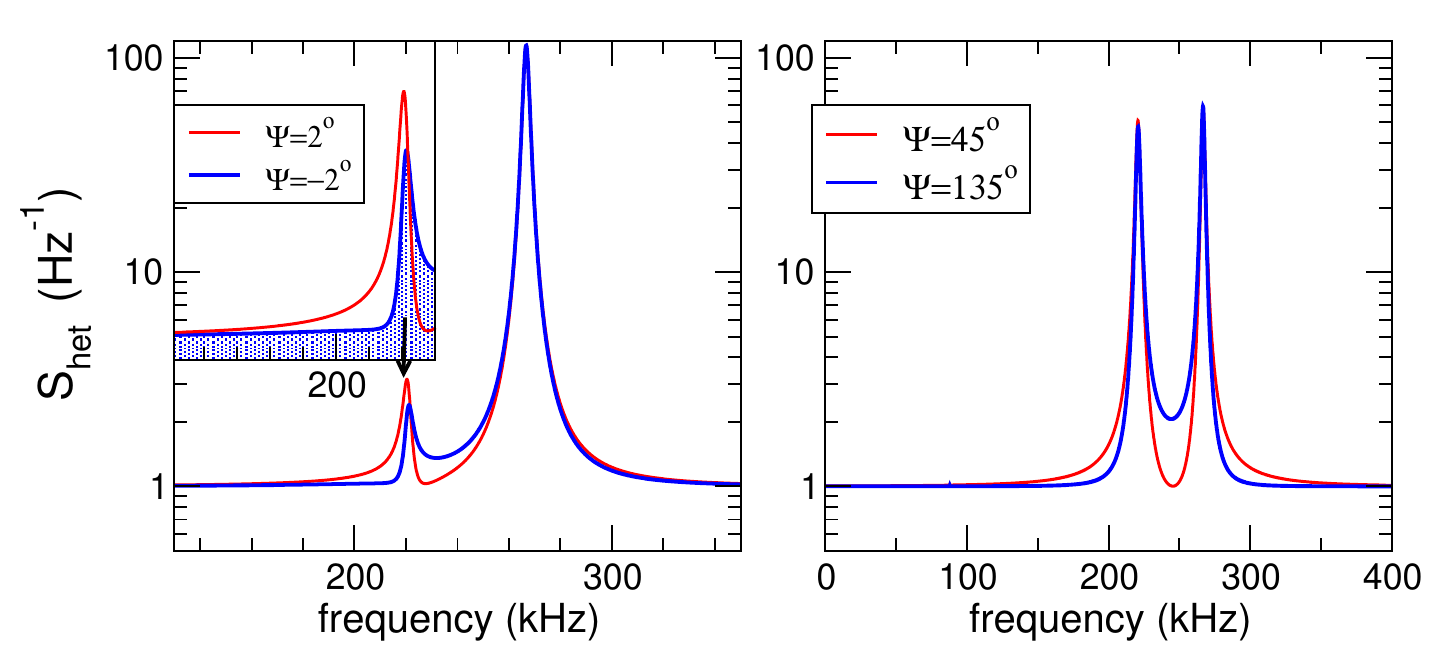}} \caption{ \textbf{(a)} For 2D quantum cooled motion in the $x-y$ plane, using cavities,  heterodyne detection gives the most sensitivity for displacement sensing.  Using parameters similar to 2D $x-y$ quantum cooling experiments \cite{Novotny2023} we show that provided $\beta  \gg 1$ the $S_{xy}$ contribution is exposed and isolated 
for $\Psi \sim 0^o$ and $\Psi \sim 90^o$.  In the left panel it is shown that the sharp asymmetry in the $y$ mode `flips' at $\Psi=0$ for the case of $\beta=10$.  The $y$ mode is not heated and remains at quantum phonon occupancy $n_y \simeq 1$.  The change in shape yields a characteristic signature of the presence and orientation of the directional force,  even in a quantum regime.  For general orientations,  the $S_{xy}$ contribution still results in a $\Psi$-dependent heterodyne PSD,  but the difference between $\Psi=\pi/4$ and $3\pi/4$ is now only quantitative, rather than a qualitative change in shape. }
\label{Fig4}
\end{figure}

Heterodyne (or homodyne) detection of the output cavity light is widely used to detect quantum-scale displacements:
\begin{equation}
S_{het} (\omega)  \approx   |\chi_c(\omega)|^2 [  g_y^2 S_{yy}   +   g_x^2 S_{xx} +
      g_xg_y  S_{xy} ] +S_{imp}
\end{equation}
where $S_{imp}$ is the imprecision noise (ideally optical quantum shot noise) and $\chi_c(\omega)$ is the optical susceptibility.  It may be seen that the heterodyne spectrum does not isolate $S_{xy}$ so its contribution is generally masked by the stronger $S_{xx}, S_{yy}$ contributions.  In addition,  there are additional artificial correlations induced by back-action induced rotations of the modes,  that for strong coupling yield dark modes 
\cite{MTTM2021, Pontin2023}.  These rotations  can be suppressed by trapping away from the cavity node \cite{Pontin2023}.  To retain the advantages of low photon heating at the node  while allowing  2D cooling implies $g_x \sim g_y \ll |\omega_x-\omega_y|$.  However this suppresses cross correlations as these require cooling widths $\Gamma_{cool} \sim  |\omega_x-\omega_y|$. 

However,  we may still detect the cross-correlation spectrum  in certain  cases.  For $g_x\sim g_y \equiv g$,  and $\beta^2\gg 1 $,  thus in the case of ultrahigh vacuum where the directed noise is dominant,  then 
$S_{het} \propto g^2(|\chi_x(\omega)|^2 \cos^2 \Psi + |\chi_y(\omega)|^2 \sin^2 \Psi  +  \text{Re}[\chi_x^*(\omega) \chi_y(\omega)] \cos \Psi \sin \Psi$.   Using parameters similar to the quantum cooling experiments in  \cite{Novotny2023},
we find that the $S_{xy}$ spectrum is exposed for $\Psi \sim 0$ for $\omega \sim  \omega_y$ and for $\Psi \sim \pi/2$ for $\omega\sim \omega_x$.  The reason for this is that e.g.,  for $\omega\sim \omega_y$,
$\frac{S_{xy}(\omega\sim \omega_y)}{S_{xx}(\omega\sim \omega_y)}= \frac{\Gamma_{cool}}{ |\omega_x-\omega_y|} \cot \Psi$.  In Fig.\ref{Fig3},  we show that even for $\frac{\Gamma_{cool}}{ |\omega_x-\omega_y|}\ll1$,
as $\cot \Psi \to \infty$ near 0, the cross-correlations are exposed.
\section{Applications}
\subsection{Temperature anisotropy}
The ability to accurately detect the presence of a directional component in the Gaussian noise baths that heat the levitated nanoparticle
opens up possibilities for thermodynamic sensing and temperature anisotropies on microscopic scales. For instance previous studies
\cite{Millen2014} have identified significant anisotropies in gas temperatures along different orientations. The present study offers the
the possibility of doing so at far more sensitive scales, as well as to identify other sources of anisotropy such as weak gas currents.

\subsection{Levitated systems and dark matter searches}

To date much  effort towards direct detection of dark matter using non-levitated platforms has been focused on Weakly Interacting Massive Particles (WIMPs), with mass ranges in the GeV to TeV range.  However,  as more parameter space for standard WIMPs is explored,  interest has  shifted towards  lower-mass candidates.  This has been accompanied by increasing interest in levitated systems as a platform for direct dark matter searches.  Levitated particles ranging from tens of nanometres to micrometers, levitated at ultrahigh vacuum  are decoupled from their environment.  They offer a high concentration of mass ($\sim 10^{10} - 10^{15}$ atomic masses) compared to atomic systems,   amplifying coupling to certain dark matter candidates.  There are a wide variety of levitation techniques:   although here we employed optical levitation,  magnetic and electrical levitation enables the trapping of heavier masses, offering potential for increased coupling,  particularly useful for bosonic ultralight dark matter searches 
\cite{DuDM2023}.  

A range of experiments have already investigated  phenomena induced by dark matter interactions. The initial focus has been on impulse searches,  considering an impulse induced by interactions via a Yukawa-like potential between a composite dark matter nugget and a levitated microsphere  \cite{dark_matter_3}.  Here,  dark matter nuggets couple to the number of neutrons in the microsphere, exemplifying the advantages of having a high concentration of mass for these types of searches.  By considering levitated nanoparticles,  the impulse sensitivity increases and allows  investigation of the products of radioactive decay.  By measuring the energy of the emitted products using the nanoparticle and external sensors, information about neutrinos can be inferred allowing searches into heavy sterile neutrinos \cite{Carney2023}.  It has been proposed that  quantum superpositions on  macroscopic scales might offer further enhancements \cite{Killian2023}.

Electrical control and trapping of levitated particles was introduced in 2015 \cite{Millen2015,fonseca2016nonlinear}: levitated particles are naturally charged with charges ranging from single to hundreds of elemental charges.  Trapping in ion traps such as Paul traps  \cite{Millen2015} can eliminate deleterious effects such as photon recoil heating observed in tweezer traps and offer an effective means of applying active cooling.  

The ability to control the net charge on a levitated system opens avenues to investigate millicharged particles bound to matter. 
Being able to neutralise a microsphere has enabled searches for charges as small as $10^{-4} e,$ \cite{Millicharges2021} improving on the state of the art by two orders of magnitude.  It also means that levitated nanospheres may provide an alternative platform to ion traps \cite{MilliCharges2023} in the search for millicharged dark matter.  Trapping of a nanoparticle in a linear Paul trap down to pressures in the $P\sim 10^{-11}$ mbar regime was recently reported  \cite{Northrup2023}. 

\subsubsection{Applications of cross-correlations}

This type of search can be augmented by considering the cross-correlations: any dark matter candidates amenable to investigation with PSDs 
can exploit the expected directionality of dark matter.  For the case of detection via PSDs in steady state, as in all proposals to exploit the heating effect of the interaction on the detection platform,  a formidable challenge is the extremely low average  mass density of dark matter $\rho_{DM}$, in the vicinity of the solar system. To provide a simple illustration, assuming
densities of 1 hydrogen atom mass per cubic cm, as a heuristic yardstick, taking pressure relation $P \simeq \rho \bar{v}^2$, and taking a streaming velocity of $ \bar{v} \sim 230 \textrm{km/s}$, corresponding to the velocity of the sun around the galaxy, would yield
an equivalent pressure of $P\sim 10^{-12}$ mbar.  

There are proposed models that envisage higher local dark matter densities: in   \cite{MilliCharges2023} a mechanism is proposed that may allow enhancements of 14 orders of magnitude. However even with more modest enhancements, the sensitivity of levitated experiments offers much promise as current levitated experiments are already approaching PSD detection at $10^{-11}$ mbar pressures in ion traps, at ambient temperatures. This simplistic comparison is not conclusive, but indicates that even non-relativistic candidates may be amenable to future detection. Higher energy impacts would be easier to detect;  cross-correlations may also be useful in the case of single recoil detections as even in that case, they may be useful to probe the directionality of the individual impact. Current experiments already achieve sensitivities corresponding to the impact from a single $\text{N}_2$ molecule at ambient temperature.

\end{document}